\documentclass{article}
\usepackage{amssymb}
\usepackage{amsmath}

\title{On tetraquark meson states}
\author{Virendra Gupta\thanks{virendra@mda.cinvestav.mx} \\Departamento de 
F\'{\i }sica Aplicada,CINVESTAV-Unidad M\'{e}rida,\\
A.P. 73 Cordemex 97310 M\'{e}rida, Yucat\'{a}n, M\'{e}xico}

\begin{document}
\maketitle
\begin{abstract} It is suggested that the narrow meson state seen in the
                 the SELEX experiment is a $(c\bar{s} s\bar{s})$
                 tetraquark state. Characteristics of other possible
                 tetraquarks formed out of $c$ and $s$ quarks and
                 antiquarks are considered.
 \end{abstract}

  SELEX experiment has reported \cite{selex} a puzzling narrow meson
state at a mass of $2632.6\pm1.6$ MeV$/c^2$ which decays preferentially to
$D^+_s(c\bar{s}) +\eta(548)$ rather than to $D^+(c\bar{d}) + K^0$ or
$D^0(c\bar{u}) +K^+$. In this note, we suggest \cite{chao} that this state is a
tetra-quark meson containing a $c$ quark and three $s$ quarks and
antiquarks. We denote this state as $D^+(c\bar{s} s\bar{s})$,
its quark content is made explicit for clarity.   

The presence of the $s\bar{s}$ pair is responsible for the observed
decay pattern.  Note that,in terms of the the SU(3) flavor
states, $\eta_1$ and $\eta_8$
 \begin{equation}  
 s\bar{s} = \frac{1}{\sqrt{3}} \eta_1 - \frac{2}{\sqrt{6}} \eta_8 ,
 \end{equation}
where,
 \begin{equation}  
  \eta_1=\frac{1}{\sqrt{3}}(u\bar{u} +d\bar{d} +s\bar{s}) \quad
  \text{and} \quad \eta_8=\frac{1}{\sqrt{6}}(u\bar{u} +d\bar{d}
  -2s\bar{s}).
 \end{equation}
  Also, we know that $\eta(548)$ is mainly $\eta_8$ with a small
  \cite{hagiwara} admixture of $\eta_1$. Consequently, it is natural
  that the main decay mode for this tetraquark is $D^+(c\bar{s}
  s\bar{s})\rightarrow D^+_s(c\bar{s}) +\eta(548)$. The $D^+ + K^0$
  and $D^0 + K^+$ modes require $d\bar{d}$ and $u\bar{u}$ pairs in the
  final state and these can appear through the 2 gluon annihilation
  process, namely $s\bar{s}\rightarrow q\bar{q}$ ($q=u,d$). Thus there
  is a natural suppression of the $D K$ modes due to the OZI-rule
  \cite{okubo}. Moreover, since gluons carry no flavor, only the the
  $SU(3)$ flavor singlet part of the $s\bar{s}$ can convert into
  $u\bar{u}$ and $d\bar{d}$ pairs. Finally,since $D^+(c\bar{s}
  s\bar{s})$ is a pure $I=0$ state it is obvious that the rates for
  decay into $D^+K^0$ and $D^0K^+$ will be equal. Furthermore,its
  decay into $ D^+_s(c\bar{s}) + \pi^0$ will be suppressed relative to
  the decay modes $D^+_s $ plus two or more pions.
  
It is interesting to consider the wave function of $D^+(c\bar{s}
 s\bar{s})$.  Since,the $\bar{s}\bar{s}$ pair is symmetric in flavor
 space, due to Pauli principle there are two possibilities for the
 color and spin wave functions of the pair assuming no relative
 orbital angular momentum: a) antisymmetric in color, symmetric in
 spin or b) symmetric in color, antisymmetric in spin.We consider both
 the possibilities.  Case(a):Here,the $cs$ pair will be in a $3^*_c$
 and this will combine with the $\bar{s}\bar{s}$ pair in a $3_c$ to
 form a color singlet tetraquark. In spin-space, $\bar{s}\bar{s}$ will
 be in total spin 1 state to satisfy the Pauli principle.However, the
 diquark cs can have total spin 0 or 1.So, ignoring any orbital
 angular momenta, the $cs$ pair will also be in a spin 1 state to give
 a tetraquark with spin 0 \cite{F1}. In this case, one can easily
 verify that this normalized spin 0 state of the tetraquark can be
 rewritten as
  \begin{equation}
       \frac{1}{\sqrt{3}}( S_{12}+ S_{21})=\frac{1}{\sqrt{3}}
       ([c\bar{s}_1]_0[s\bar{s}_2]_0 + [c\bar{s}_2]_0[s\bar{s}_1]_0).
  \end{equation}
 . In this form the symmetry between the two strange antiquarks
  denoted as $\bar{s}_1$ and $\bar{s}_2$ is explicit. The notation is
  that $[qq']_0$ means $q$ and $q'$ are in a normalized total spin 0
  state.For notational convenience,we have defined
  $S_{ij}=[c\bar{s}_i]_0 [s\bar{s}_j]_0$ for (i,j)=(1,2)or(2,1) in
  which $\bar{s}_i$ and $\bar{s}_j$ form a spin singlet with c and s
  respectively.The $S_{12}$ and $S_{21}$ are normalized to unity but
  $<S_{12}|S_{21}>=1/2 $.Further,in color space,define the
  wavefunctions $C_{ij}=[c\bar{s}_i]_{1_c} [s\bar{s}_j]_{1_c}$ for
  (i,j)=(1,2)or(2,1) in which $\bar{s}_i$ and $\bar{s}_j$ form a color
  singlet with c and s respectively.The notation used is that
  $[qq']_{1_c}$ means $q$ and $q'$ are in a normalized color singlet
  ($1_c$) state.  Consequently,$C_{12}$ and $C_{21}$ are normalized to
  unity and $<C_{12}|C_{21}> =1/3$.The normalized color- spin
  wavefunction ,for this case is
  \begin{equation}
           \psi_a =\frac{1}{2}(C_{12}- C_{21})(S_{12}+S_{21}).
   \end{equation}
    Case(b):In this case,the color wavefunction is symmetric, $6^*_c$ for
    the $\bar{s}\bar{s}$ pair and $6_c$ for the cs pair,while the spin
    wavefunctions are antisymmetric,both the pairs being in a total
    spin 0 state.In the above notation,the normalized color-spin wavefunction
    is simply
   \begin{equation}
          \psi_b =\sqrt{\frac{3}{8}}(C_{12}+ C_{21})(S_{12}-S_{21}).
   \end{equation}
    The symmetry/antisymmetry between the two strange antiquarks with
    respect to color and spin is explicit.Both the wavefunctions are
    overall antisymmetric under the interchange of the two antiquarks.
    The spin and color structure of the wave function in both the
    cases favours the $D^+_s \eta$ decay mode.In general,the
    wavefuncion of the tetraquark would be a linear combination of the
    two possible cases,namely
    \begin{equation}
         \psi= [\cos\phi\, \psi_a +\sin\phi\, \psi_b]~,
     \end{equation}
     if the color-spin forces are attractive in both the cases.
     The mixing angle $\phi$ will be determined by  the nature of
    the color-spin forces in the tetraquark.Its value is important for
    the decay width of the tetraquark as discussed later.  Given our
    assumptions the $D^+(2632)$, if it is the tetraquark $
    D^+(c\bar{s} s\bar{s})$, should have $J^{P}=0^{+}$.

  One can make a rough estimate of g, the  dimensionless effective strength of
  the decay vertex for  $D^+(2632)\rightarrow D^+_s(cs^-) +\eta(548)$.
  For two-body decay at rest,the width
  \begin{equation}
                   \Gamma=\frac{1}{8\pi}\frac{k}{m^2_0}|A|^2,
  \end{equation}
   where $m_0$ is the mass of $D^+(2632)$, $k$ is the momentum of a
   decay particle and $A$ is the transition amplitude.In our case
   $k$=326 MeV.  For $s$-wave decay we take $A=g m_0 a(\theta_P)$,
   where $a(\theta_P)$ determines the amplitude for $\eta$ in $s\bar
   s$ and $\theta_P$ is the mixing angle between $\eta_1$ and
   $\eta_8$.In fact,
  \begin{equation}
       s\bar s= - a(\theta_P) \eta(548) + b(\theta_P)\eta'(958), 
 \end{equation}
   where $a(\theta_P)=\frac{1}{\sqrt{3}}[\sqrt{2}\cos\theta_P +
   \sin\theta_P]$ and $b^2=1- a^2$. Using $\theta_P= -10.1^o$
   \cite{hagiwara}, $a(\theta_P)$=0.7 \cite{F2}. Since, experimentally
   $\Gamma < $17 MeV we obtain $g^2(m_0)/(4\pi)<$ 0.21. This value at
   $m_0$=2632 MeV is not unreasonably small considering that the QCD
   couplant $\alpha_s(3\text{GeV})\sim 0.25-0.3$ \cite{orbit}.This
   estimate does not take into account the fact that
   the $\bar{s}\bar{s}$ pair is in an antisymmetric state in the
   tetraquark.Due to this,the amplitudes for decay into the final
   states $D^+_s(c\bar{s}_1) + \eta(s\bar{s}_2)$ and
   $D^+_s(c\bar{s}_2) + \eta(s\bar{s}_1)$ will have opposite
   signs.Consequently,only the antisymmetric part of the final state
   wavefunction,namely $[C_{12}S_{12} -C_{21}S_{21}]$,in the above
   notation,will contribute giving a decay amplitude proportional to
   the factor
   \begin{equation}
    F(\phi)=<\psi|C_{12}S_{12} -C_{21}S_{21}>
    =\cos\phi+\sqrt{\frac{2}{3}}\sin\phi
 \end{equation}
    This shows that the width for a tetraquark in a pure $\psi_b$ state
   ($\phi=\frac{\pi}{2}$) will be 2/3 times smaller than that for
   a tetraquark in a pure $\psi_a$ state ($\phi=0$).It is interesting
   to note that for a negative $\phi$ this factor would inhibit the
    $D^+_s \eta$ decay mode\cite{F3}. If so, this may be responsible for the
   observed narrow width.For example,for $\phi=-(20^o-30^o )$,one
   obtains $F(\phi)= 0.66-0.46$ implying a reduction in the width by a
   factor of 0.44 to 0.21 In fact,for $\tan\phi=-\sqrt{3/2}$ (that is,
   $\phi=-50.77^o$), $F(\phi)$ vanishes and this decay mode would be absent!

  The OZI-suppression factor $f$ for the $D^0K^+ $ decay amplitude
  (for which $k=500$ MeV) can be estimated using the experimental
  ratio $0.16\pm0.06$ for the two modes
  $\cite{selex}$. \emph{Assuming} $g$ \emph{to be the same for the two
  modes,} one obtains $f=0.226$ which is plausible because the color
  singlet $s\bar{s}$ can convert into a $u\bar{u}$ through two or more
  gluons, implying $f\sim\alpha^2_s$ or one expects that $f\sim
  0.1$.So,the $D^+_s \eta$ decay mode will still be favoured if the
  value of $\phi$ is say at least $\pm10^o$ away from -50.77$^o$
  giving $F\sim0.21$ or so.

   The determination of the mixing angle $\phi$ and
  the width requires a dynamical calculation which would become
  necessary after the parameters of the SELEX state are confirmed in
  the future.

  We now speculate about other similar tetraquark states. Of the many
  possible tetraquarks we limit our discussion to mesons containing
  only $c$ and $s$ quarks and antiquarks because these are likely to
  have distinctive decay modes. One can form 6 tetraquarks including
  $D^+(c\bar{s} s\bar{s})$. For clarity we make the quark content of
  the state explicit. For the discussion below we assume that all
  orbital angular momenta are absent. As a result,the wavefunction
  discussion applies \textit{ mutatis mutandis} to all these
  tetraquarks except for one(see no.4 below).

 1.Tetraquark $ D^0(c\bar{c}c\bar{c})$. The 4 charmed quarks will have
 a symmetric spin wave function namely, $[c(1)\bar{c}(1)]_0
 [c(2)\bar{c}(2)]_0 +[c(1)\bar{c}(2)]_0 [c(2)\bar{c}(1)]_0$ up to a
 normalization factor.Since,with no orbital angular momentum, $
 [c\bar{c}]_0$ has $ J^{PC}=0^{-+}$,the tetraquark will have $
 J^{PC}=0^{++}$.  energetically allowed, the favoured decay mode would
 be
 \begin{equation}
       D^0(c\bar{c}c\bar{c}) \rightarrow  \eta_c(c\bar{c}) + \eta_c(c\bar{c}).
  \end{equation}
   However, if this decay mode is energetically forbidden, that is,the
    (mass of $D^0(c\bar{c}c\bar{c}))$ $=m_1 <5960$ MeV, then
    OZI-suppressed decays due to the annihilation of one $ c\bar{c}$
    pair into a $q\bar{q}$ pair($q=u,d$ or $s$) would dominate. The
    typical hadronic modes expected are $D^+_s + D^-_s, D^+ + D^-$ or
    $D^0 + \bar{D}^0$ if $ m_1> 4000$ MeV. In this case,one would
    guess that the tetraquark would have a width of the order of the
    width of $ \eta_c(c\bar{c})$ which is 13 MeV .

 2.Tetraquark $D^+(c\bar{c}c\bar{s})$. The arguments for a symmetric
   spin wave function apply because of the $cc$ pair and it can be
   obtained from Eq(3) by appropriate changes in the quark
   flavours. Thus, for this $I=0$ and $J^P=0^+ $ meson,the
   $(c\bar{c})$ and $(c\bar{s})$ pairs have zero total spin. One
   expects the dominant mode to be
  \begin{equation}
    D^+(c\bar{c}c\bar{s}) \rightarrow  \eta_c(c\bar{c}) + D^+_s(c\bar{s}),
   \end{equation}
   provided that the (mass of$ D^0(c\bar{c}c\bar{s})))=m_2>4948 $
   MeV. If $m_2$ is smaller, then its OZI-suppressed modes (resulting
   from $ c\bar{c}\rightarrow q\bar{q}$) should indicate its
   presence. The likely hadronic modes are $ D^+_s +\eta(548), D^0
   +K^+$ or $D^+ +K^0 $ if $m_2>2600 $ MeV. In this case, the
   tetraquark would probably have a width similar to that of $\eta_c$.

   3.Tetraquark $D^{++}(c\bar{s}c\bar{s})$. The flavour symmetric $cc$
     (or $\bar{s}\bar{s}$) implies that both the $c\bar{s}$ pairs are
     in a spin zero state. Consequently, the likely dominant decay is
     expected to be
     \begin{equation}
              D^{++}(c\bar{s}c\bar{s}) \rightarrow D^+_s + D^+_s,
   \end{equation}
    provided that the ( mass of $ D^{++}(c\bar{s}c\bar{s}))=m_3 >3940$
    MeV.  The possibility of (baryon +antibaryon) decay modes in this
    case (or for the tetraquarks in cases 1 and 2 above) requires a
    much higher mass. Since there is no $q\bar{q}$ which can
    annihilate into gluons, consequently, in this case there are no
    OZI-suppressed hadronic decay modes. This case is very interesting
    because if $m_3<3940$MeV it can decay only through weak
    interactions! Note that the exchange mechanism is absent here. So,
    the weak decays can take place through the spectator mechanism(
    eg.$ c \rightarrow W^+ + q$,with $q=d$ or $s$) or through the
    annihilation mechanism, namely $c + \bar{s} \rightarrow W^+$. The
    latter gives lighter final states, namely $D^+_s + l^+ + \nu_l
    $(l=e or $ \mu)$). The e ($ \mu$) modes are allowed if
    $m_3>1970(2074)$ MeV. In contrast the spectator mechanism cangive
    light final states like $D^+_s + K^0 +l^+ + \nu_l $ and $ D^+_s
    +K^+ +\pi^0$,which require $m_3> 2463$ and 2603 MeV.  Guessing
    that $m_3 $ is approximately 3,000 Mev these decays would be a
    striking signature ofthis tetraquark. Further,its quantum numbers
    are unusual.  It is an $I=0$ state which is doubly charged,charmed
    and strange!

  4.Tetraquark $D^0(c\bar{c}s\bar{s})$. This contains 2 charmed and 2
   strange quarks like the tetraquark in case 3 above but unlike it
   carries no charge, charm and strange\-ness. So, one might expect
   its mass $m_4$ to be near $m_3$. For this case,there is no Pauli
   principle argument for the symmetry of the spin wave
   function. Possible likely decay modes are
    \begin{align}
            D^0(c\bar{c}s\bar{s}) &\rightarrow \eta + \eta_c,
                                  &\text{if}\quad m_4>3528 MeV\\
                                  &\rightarrow \eta + \ J/\psi,
                                  &\text{if}\quad m_4>3645 MeV\\
                                  &\rightarrow D^+_s + D^-_s,
                                  &\text{if}\quad m_4>3940 MeV
      \end{align}
        These decays, if energetically possible can determine the $
     J^{PC}$ of this meson. For instance, if $J^{PC}= 0^{++}(0^{--})$
     then the decay in Eq(8) is forbidden(allowed) while the other two
     are allowed(forbidden). In case these decays are not allowed
     energetically, then the OZI-suppressed decays due to annihilation
     of $ c\bar{c}$ into $ q\bar{q}$ ($q=u,d,s$) have to be
     considered. In that case the width of $D^0(c\bar{c}s\bar{s})$
     would be comparable to that of $\eta_c$. Possible decay modes,
     which come to mind, if the tetraquark has $J^{PC}=0^{++}$, are $K
     + \bar{K},\eta + \eta $ if 1100 MeV$<m_4<$3500 MeV.
    
     5.Tetraquark $D^+(c\bar{s}s\bar{s})$.This has been discussed
       above to be a canditate for the D(2632) meson observed in the
       SELEX experiment.

     6. Tetraquark $S^0(s\bar{s}s\bar{s})$.  This should be the
       lightest of the six tetraquarks consideed here. Pauli principle
       argument applies in this case and each $s\bar{s}$ will be in
       total spin zero state. The favoured decay mode is expcted to be
       $\eta +\eta$. Interestingly enough there is a possible
       candidate for $S^0$, namely $f_0(1370)$ \cite{hagiwara} which
       was observed explicitly as a $\eta\eta$ resonance in the
       reaction $p\bar{p} \rightarrow \pi^0\eta\eta$ \cite{amsler}.
       In other experiments \cite{scalar}, the $f_0(1370)$ interferes
       with the $f_0(1500)$. Particle Data Group\cite{hagiwara}give
       the mass and width of the $f_0(1370)$ as 1200 to 1500 MeV and
       200 to 500 MeV. These are well defined for $f_0(1500)$, namely
       $1500\pm 10$ MeV and $112 \pm 10$ MeV(could this be the elusive
       $S^0$??). Both the $f_0$'s decay into $\eta\eta$,$K\bar{K}$ and
       $\pi\pi$. For $S^0$, $K\bar{K}$ decays are OZI-suppressed
       because one $s\bar{s}$ pair must annihilate through gluons into
       $ q\bar{q}$ ($q=u,d$) while, the $\pi\pi$ decays are doubly
       suppressed since both the $s\bar{s}$ pairs must annihilate into
       $ q\bar{q}$.At present the branching fractions are not known
       experimentally. We note that in general the OZI- suppression
       depends on the particular case because the quark-gluon coupling
       strength ($\alpha_s$) decreases with the energy involved in the
       process. Consequently,the OZI-suppression of the decay modes
       arising from $c\bar{c} \rightarrow q\bar{q}$ will be much
       stronger than those due to $s\bar{s} \rightarrow q\bar{q}$. For
       $S^0$,the OZI-suppression of the $\pi\pi$ modes relative to the
       $K\bar{K}$ modes may be compensated by the large phase space
       available to the former modes.  The whole situation is rather
       complicated and unclear. One has to await more experimental
       data.
      
      In this note a qualitative discussion of tetraquark mesons
      formed out of $c$ and $s$ quarks and antiquarks has been
      presented. Distinctive decay characteristics of these states are
      pointed out which may help to identify them experimentally. Some
      guess for their masses would be helpful, however ,in case of
      mesons one can only make uneducated guesses!

\section*{Acknowledgement} It is a pleasure to thank Antonio Bouzas for
discussions and help.


\begin{thebibliography}{9}
\bibitem{F1} The possibility of spin 0 for the cs diquark is ruled
  out because,in this case.it will give a tetraquark with $J^P=1^+$ and
  this cannot decay into two pseudoscalars.However,if we allow one unit
  of angular momentum, the cs pair will be in a $1^-$ state to give a
  tetraquark with $J^P=0^-,1^-$ and $ 2^-$.Of these only the $1^-$ can
  decay into two pseudoscalars.A spin 1 cs diquark will form a
  tetraquark with $J^P=0^+,1^+$ and $ 2^+$.The first case is considered
  in the text.Note, that a $2^+$ tetraquark can decay into two
  pseudoscalars with l=2 in the final state.For $J^P=1^- or 2^+$,the
  observed narrow width would not be a problem.See text.
\bibitem{F2} The value of $\theta_P$ is sensitive to a small
  breaking,$\Delta$,of the Gell-Mann-Okubo mass relation.A positive 
  $\Delta$  decreases $a(\theta_P)$.
\bibitem{F3} The opposite can happen for positive $\phi$.$F(\phi)$ incresaes
  from 1($\phi=0$) to a maximum of 1.29 for $\phi=39.23^o$ and then
   decreases to 0.81 for $\phi=90^o$.
\bibitem{orbit} Of course, if $D^+(2632)$ has $J^P=1^-,2^+\ldots$,
   then amplitude $A$ would be supressed by $(k/m_0)^l$, $l=
   1,2,\ldots$ Since, $k/m_0$=0.124, this would not give an useful
   upper limit on $g$, also it would imply a narrow width.
\end{thebibliography}
\end{document}